\documentclass[10pt]{article}
\usepackage{amsmath}
\usepackage{amsfonts}
\usepackage{graphicx}
\usepackage{cite}
\usepackage{epstopdf}
\usepackage{url}
\usepackage{booktabs}
\usepackage{array}
\usepackage{setspace}
\newcolumntype{x}[1]{%
>{\centering\hspace{0pt}}p{#1}}%

\newcommand{\refeq}   [1] {eq.~(\ref{#1})}
\newcommand{\refeqt}  [2] {eqs.~(\ref{#1})~and~(\ref{#2})}
\newcommand{\refeqn}  [2] {eqs.~(\ref{#1})~-~(\ref{#2})}

\begin{document}

\noindent
\large{\bfseries{Delivery of time-varying stimuli using ChR2}} \vspace{0.5cm} \\
Tatjana Tchumatchenko$^{1,2,*}$, Jonathan P. Newman$^{3,*}$, Ming-fai Fong$^{3,4}$, Steve M. Potter$^{3}$ \vspace{0.1cm}\\
$^{1}$ Center for theoretical Neuroscience, \\
Columbia University College of Physicians and Surgeons, \\
1051 Riverside Dr, New York, NY 10032, USA \\
$^{2}$ Max Planck Institute for Brain Research, \\
Deutschordenstr. 46, 60528 Frankfurt am Main, Germany \\
$^{3}$ Dept. of Biomedical Engineering, Georgia Institute of Technology, 313 Ferst Drive, Atlanta, GA, 30332, USA\\
$^{4}$ Dept. of Physiology, Emory University School of Medicine, 201 Dowman Drive. Atlanta, GA 30322, USA\\
$*$ Co-first authors; These authors contributed equally.\\

\noindent
{\bf Acknowledgments:} We thank L. Abbott, K. Miller and S. Fusi for fruitful discussions. We thank M. LaPlaca for providing tissue and J.T. Shoemaker for performing tissue harvests. We thank N. Laxpati, and the Kaplan Lab for assistance with virus production. We thank P. Wenner for patch recording expertise. T.T. is funded by Volkswagen Foundation and the Center for Theoretical Neuroscience, Columbia University. Experimental work was supported by NSF COPN grant 1238097 and NIH grant 1R01NS079757-01. Additionally, this work was supported by NSF GRFP Fellowship 08-593 to J.P.N., NSF GRFP Fellowship 09-603 to M.F., and NSF IGERT Fellowship DGE-0333411 to J.P.N and M.F. 

\vspace{0.5cm}
\noindent
{\bf Date:} 2013.01.14

\newpage
\setcounter{page}{1}
\doublespacing
\section*{Abstract}
To understand sensory processing in neuronal populations, it is necessary to deliver stimuli to the sensory organs of animals and record evoked population activity downstream. However, the pathways from sensory input to synaptic currents in cells that are several synapses removed from sensory organs are complex. Intrinsic noise and uncontrolled modulatory input from other brain regions can interfere with the delivery of well-defined stimuli. Here we investigate the ability of channelrhodopsins to deliver precise time-varying currents to neurons at any point along the sensory-motor pathway. To do this, we first deduce the amplitude response function of channelrhodopsin-2 (ChR2) using a three state Markov model of channel kinetics. With biophysically realistic parameters, this function supports a relatively broad signal passband and contains a resonance. We confirm the validity of our predicted amplitude response function using time-varying optical stimulation of excitatory neurons that express either wild type ChR2 or the ChR2(H134R) mutant. Together, our results indicate that ChR2-derived optogenetic tools are useful for delivering repeatable, time-varying currents to genetically-specified populations over a physiologically-relevant frequency band.
 
\section*{Introduction}
\label{sec:Introduction}
\noindent
The network response to time-varying input is at the core of cognitive and sensory processing. To understand how neuronal networks process time-varying input, precisely defined stimuli are delivered via a sensory organ, and evoked spiking activity is recorded downstream. The power of this technique for deducing network encoding properties has been demonstrated in a number of sensory preparations, for instance, the retina and olfactory system~\cite{Warland1997,Chichilnisky2001,Geffen2009}. However, as stimuli delivered to sensory organs propagate to higher brain areas, intrinsic noise and modulatory input from secondary brain regions can interfere with controlled input signals. For studies that seek to understand the function of neural circuits that are several synapses removed from sensory input, the direct introduction of time-varying currents to neural population may allow a more straightforward deduction of the circuit response properties.

Optogenetic methods allow precise control of spike times using brief light pulses to excite light-gated cation channels, such as channelrhodopsin-2 (ChR2)~\cite{Mattis2011,Gunaydin2010}. Pulsed stimulation reliably dictates a spiking response that is tightly locked with each stimulus by briefly overriding neuronal dynamics. It stands in contrast to the highly variable, sub-threshold currents recorded from cortical neurons during natural sensory processing {\it in vivo}~\cite{destexhe2003}. We hypothesized that using relatively low intensity, continuously modulated optical stimuli to excite ChR2 might allow conductance fluctuations that imitate natural synaptic bombardment {\it in vivo} and leave the decision of when to spike to individual cells~\cite{Mainen1995,TchumJN}. Surprisingly, while the response properties of microbial opsins to optical pulses have been studied extensively~\cite{Mattis2011}, little is known about their ability to relay fluctuating light signals.

In order for ChR2 to be useful for delivering time-varying currents, it must allow (1) sufficient bandwidth to mimic synaptic communication and (2) repeatable current waveforms to be delivered across trials. Here, we address these requirements theoretically and experimentally. We find that wild-type ChR2 (wtChR2)~\cite{Boyden2005} supports significant photocurrents from $\sim$0-69 Hz and H134 mutant (ChR2(H134R)~\cite{Nagel2005} from $\sim$0-37 Hz. We show that evoked current waveforms are extremely repeatable across trials. Finally, we find that wild type ChR2 supports a strong resonance with a natural frequency around 10 Hz. This resonance is present, but not significant in the H134R mutant.

\section*{Materials and Methods}
\subsection*{Derivation of ChR2's frequency response function}
\noindent
To derive ChR2's light to photocurrent response function, we considered a three state Markov model of ChR2's state transition kinetics, which was introduced in~\cite{Nagel2003} and is depicted in Fig.~\ref{fig:ResponseChRVsP}(a). The rate equations governing channel transitions are
\begin{align}
\frac{d O(t)}{dt}&=\epsilon\phi(t) C(t)-\Gamma_d O(t) \label{eq:DiffEqOpenProb1}\\
\frac{d D(t)}{dt}&=\Gamma_d O(t)-\Gamma_r D(t) \label{eq:DiffEqOpenProb2}\\
C(t) &= 1-O(t)-D(t), \label{eq:DiffEqOpenProb3}
\end{align}
where the state variables $O(t)$, $D(t)$, and $C(t)$ are the probabilities of a channel being open, desensitized, or closed, respectively. $\Gamma_d$ and $\Gamma_r$ are the rates of channel desensitization and recovery. $\epsilon$ is the quantum efficiency of ChR2 and $\phi(t)$ is the instantaneous photon flux for a single channel. $\phi(t)$ can be modulated by changing the brightness of the optical stimulator. The evoked photocurrent across the cell membrane is proportional to the fraction of open channels. Therefore, under the assumption that a cell expresses a large number of channels, identifying ChR2's frequency response is equivalent to finding the frequency response of $O$ in the continuum limit. We refer to ChR2's frequency response function as $F_{\mathrm{ChR2}}(\omega)$.

$F_{\mathrm{ChR2}}(\omega)$ can be obtained by first considering a small sinusoidal light signal with a constant offset $\phi_0$,
\begin{align}
\phi(t)=\phi_0+\delta\phi \exp(j \omega t), 
\end{align}
where $\omega=2 \pi f$ and $f$ is the frequency of the sinusoid in Hz and $j=\sqrt{-1}$. The first order dynamics of the open and closed probabilities can then be described by a constant offset and small periodic component,
\begin{align}
O(t)&=O_0+\delta O_0 \exp(j \omega t) \label{eq:LinMod1} \\
D(t)&=D_0+\delta D_0 \exp(j \omega t).\label{eq:LinMod2}
\end{align}
In the linear regime, changes in the open state, $\delta O$, or the desensitized state, $\delta D$, are proportional to changes in optical input, $\epsilon\delta \phi$. The proportionality factors for the open and desensitized states are the frequency response functions $F_{\mathrm{ChR2}}(\omega)$ and $G_{\mathrm{ChR2}}(\omega)$, respectively,
\begin{align}
\delta O&= \epsilon\delta\phi F_{\mathrm{ChR2}}(\omega)\label{eq:DeltaO} \\
\delta D&=\epsilon\delta\phi G_{\mathrm{ChR2}}(\omega)\label{eq:DeltaD}.
\end{align}
Differentiating \refeqt{eq:LinMod1}{eq:LinMod2} and inserting the result into \refeqt{eq:DiffEqOpenProb1}{eq:DiffEqOpenProb2} leads to
\begin{align}
j\omega\delta O\exp(j\omega t)= &\left[ \epsilon \phi_0 (1-O_o-D_o) \right] + \Big[ \epsilon \delta \phi (1-O_o-D_o) \label{eq:LinMarkov1} \\  
 &+ \epsilon \phi_0 (\delta O - \delta D) -\Gamma_d \delta O\Big]\exp(j\omega t) + \mathcal{O}(2) \nonumber \\
j\omega \delta D\exp(j\omega t)=&\left[ \Gamma_d O_0 - \Gamma_r D_0 \right] + \left[  \Gamma_d \delta O - \Gamma_r \delta D \right]\exp(j\omega t).\label{eq:LinMarkov2}
\end{align}
By dropping all but the first-order terms of \refeqt{eq:LinMarkov1}{eq:LinMarkov2} (meaning those terms proportional to $\exp(j\omega t)$), and removing the common factor $\exp(j\omega t)$, linear changes in the open and desensitized states due to changes in light power are given by
\begin{align}
j\omega\delta O&=\epsilon (1-O_0-D_0) \delta \phi +(\epsilon\phi-\Gamma_d)\delta O-\epsilon\phi\delta D \\
j\omega \delta D&=\Gamma_d \delta O-\Gamma_r \delta D,
\end{align}
where $(1-O_0-D_0) = C_0$ is the steady-state probability of the channel being closed. Performing the necessary substitutions to solve for $\delta O$ results in
\begin{align}
\delta O&=\epsilon\delta \phi \left[ \frac{C_0(j\omega +\Gamma_r)}{-\omega^2 +j\omega(\Gamma_r+\epsilon\phi+\Gamma d)+\epsilon\phi\Gamma_r+\epsilon\phi\Gamma_d+\Gamma_r\Gamma_d}\right]  \label{eq:LinChange1}.
\end{align}
Finally, referencing \refeq{eq:DeltaO}, ChR2's frequency response function is calculated by dividing the left hand side of \refeq{eq:LinChange1} by $\epsilon \delta \phi$, 
\begin{align}
F_{ChR2}(\omega)&= \frac{\delta O}{\epsilon\delta\phi} =\frac{C_0(j\omega +\Gamma_r)}{-\omega^2 +j\omega(\Gamma_r+\epsilon\phi+\Gamma_d)+\epsilon\phi\Gamma_r+\epsilon\phi\Gamma_d+\Gamma_r\Gamma_d},\label{eq:ComplexR}
\end{align}
and the amplitude response is then given by
\begin{align}
|F_{\mathrm{ChR2}}|&=\frac{C_0\sqrt{\omega^2 +\Gamma_r^2}}{\sqrt{(-\omega^2+\epsilon\phi\Gamma_r+\epsilon\phi\Gamma_d+\Gamma_r\Gamma_d)^2 +(\omega(\Gamma_r+\epsilon\phi+\Gamma_d))^2}}.\label{eq:ChRResponseAmplitude}
\end{align}
In the high frequency limit, \refeq{eq:ComplexR} reduces to $\frac{C_0(j\omega)}{-\omega^2 +j\omega(\Gamma_r+\epsilon\phi+\Gamma_d)}\propto\frac{C_0}{j\omega/(\Gamma_r+\epsilon\phi+\Gamma_d) +1}$. This indicates that the frequency cutoff of  $F_{\mathrm{ChR2}}(\omega)$ is determined by a low-pass filter with a time constant $1/(\Gamma_r+\epsilon\phi+\Gamma_d)$. Thus increasing the baseline light intensity or the transition rates will result in a broader frequency range over which $F_{\mathrm{ChR2}}(\omega)$ supports significant photocurrents.

\subsection*{Experimental methods}
\label{sec:LinearResponseestimate}

{\bf Culturing methods} Our culturing methods are described in detail elsewhere~\cite{Hales2010}. All experiments were carried out in accordance with the U.S. Public Health Service's Policy on Humane Care and Use of Laboratory Animals and the Guide for the Care and Use of Laboratory Animals using a protocol approved by the Georgia Tech IACUC. Timed-pregnant female rats were anesthetized with inhaled isoflurane and killed by decapitation. Whole brains were excised from embryonic day 18 (E18) rats. Cortical tissue was digested in a solution of 20 U$\cdot$ml$^{-1}$ papain (Sigma-aldrich). Following enzymatic digestion, cells were mechanically dissociated using 3 to 5 trituration passes through a p1000 pipette tip. The resulting cell suspension was filtered through with a 40 $\mu$m cell strainer and then centrifuged at 200$\cdot g$ to remove large and small debris, respectively. The cell pellet was diluted to 2500 cells$\cdot\mu$L$^{-1}$. Approximately 50,000 cells in a 20 $\mu$L drop were plated at onto a $\sim$2 mm diameter area on glass bottom petri dishes, resulting in $\sim$2,500 cells$\cdot$mm$^{-2}$ on the culturing surface. $0.75$ mL of the culturing medium was exchanged every three days, for each culture. Cultures dishes were sealed with a Teflon membrane~\cite{Potter2001} and stored in an incubator regulated to $35\,^{\circ}\mathrm{C}$, $5\%$ $\mathrm{CO_2}$, $65\%$ relative humidity. 

{\bf ChR2 expression system} AAV2-CaMKll$\alpha$::hChR2(H134R)-mCherry at 4$\cdot10^{12}$ c.f.u.$\cdot$ml$^{-1}$ was produced by the University of North Carolina at Chapel Hill Vector Core. AAV2-CaMKll$\alpha$::wtChR2-mCherry at 4$\cdot10^{12}$ c.f.u.$\cdot$ml$^{-1}$ was produced by the Kaplitt lab (Cornell University) using plasmid DNA for CaMKII$\alpha$::wtChR2-mCherry obtained from the K. Deisseroth (Standford University). At 1 to 5 days {\it in vitro} (DIV), viral aliquots were diluted to 1$\cdot10^{12}$ c.f.u.$\cdot$ml$^{-1}$ using glial-conditioned culturing medium. 1 $\mu$L of diluted viral solution was added to 1 mL culturing medium for a final infection concentration of 1$\cdot10^9$ c.f.u.$\cdot$ml$^{-1}$. Cultures were then incubated for 3 days before the culturing medium was exchanged. The fluorescent signal of the mCherry reporter protein was monitored in 3 sister cultures over the days post infection, and it increased monotonically before plateauing at $\sim$3 weeks {\it in vitro}. All experiments were carried out on cultures that were 3 to 4 weeks old.

{\bf Optical stimulation} A 10-watt light emitting diode (LED) was used for optical stimulation, with peak emission wavelength of 465 nm and $\sim$20 nm full width at half maximum intensity (LZ4-00B200, LEDEngin, San Jose, CA). To deliver optical stimuli to cultured neurons, the LED was focused into the epi-illumination port of an E600FN upright microscope (Nikon Corporation, Tokyo, Japan) and passed through a 40X objective lens. The light power produced by LEDs is affected by their temperature. Additionally, the relationship between forward diode current and irradiance is a static non-linearity. To compensate for these factors and deliver distortion-free optical stimuli, we designed a precision current source with integrated optical-feedback to drive our LED (Fig.~\ref{fig:Reliability}(a)). This circuit measures the instantaneous optical power produced by the LED using an amplified photodiode. It then adjusts the current sourced to the LED such that the optical power measurement matches a reference voltage supplied by a digital to analog converter (DAC; LIH 1600, HEKA Electronik, Lambrecht/Pfalz, Germany). The circuit can precisely modulate the LED brightness over a bandwidth of $~90$ kHz (Fig.~\ref{fig:Reliability}(b-c)). A full design specification for the device is available online\footnote{\url{https://potterlab.gatech.edu/newman/wiki}}.

{\bf Intracellular recordings} Whole-cell voltage-clamp recordings were conducted on neurons expressing the mCherry reporter protein. Recordings were performed in a continuous perfusion of artificial cerebrospinal fluid (aSCF) bubbled with 95\% O$_{2}$ and 5\% CO$_{2}$ to maintain a pH of 7.4. The aSCF solution contained (in mM) 126 NaCl, 3 KCl, 2 CaCl$_{2}$, 1 NaH$_2$PO$_{4}$, 25 NaHCO$_{3}$, 1.5 MgSO$_{4}$ and 25 D-glucose. The temperature of the extracellular medium was regulated to $35\,^{\circ}\mathrm{C}$ using an inline heater (Warner Instruments, Hamden, CT). 1.5 mm outer diameter, 1.1 mm inner diameter borosilicate glass capilaries (Sutter Instruments, Novato, CA) were pulled into patch pipettes and filled with a solution containing (in mM) 100 K-gluconate, 30 KCl, 3 ATP, 2 MgSO$_{4}$, 0.5 ethylene glycol tetraacetic acid and 10 HEPES adjusted to pH 7.4 using 0.1 M KOH. Filled pipettes had resistances of 4-8 M$\Omega$. All recordings were performed using HEKA EPC8 amplifier and PatchMaster control software in whole-cell voltage clamp mode. Cells were held at -70 mV and membrane current measurements were amplified and low-pass filtered at 3 kHz before being digitized at 20 kHz and streamed to disk. Access resistance and seal resistance were monitored between stimulation protocols. All experiments were performed in the presence of 40 $\mu$M 6-cyano-7-nitroquinoxaline-2,3-dione (CNQX), 50 $\mu$M (2R)-amino-5-phosphonovaleric acid (AP5), 20 $\mu$M bicuculline in order to prevent synaptic transmission. Whole-cell recordings were analyzed offline in MATLAB (The MathWorks, Natick, MA). 

\subsection*{Experimental verification of frequency response functions}
To estimate the frequency response of ChR2, we delivered optical stimuli, $s(t)$, consisting of $T$=10 second realizations of a Gaussian (Ornstein-Uhlenbeck) noise process while recording evoked intracellular currents, $I_i(t)$, within a single cell, $i$. $s(t)$ was generated using
\begin{align}
s(t_{n+1}) = \mu_s + s(t_n)\exp(-dt/\tau_s) + \sigma_s\xi(t_n)\sqrt{1-\exp(-2dt/\tau_s)},
\end{align}
where $s(t_1)=0$, $\mu_s=$ 0.4, and $\sigma_s$= 0.08 mW$\cdot$mm$^{-2}$ are the initial condition, mean, and standard deviation of the process respectively, $\tau_s=50$ ms is the correlation time $s(t)$, $dt=40$ $\mu$s is the DAC update period, and $\xi(t_n)$ is a random variable drawn from the standard normal distribution. Each cell was exposed to the a single, repeated realization of $s(t)$ for $k=10$ trials. The first second of each trial was ignored to remove the non-stationary effects of the stimulator turning on. The recorded intracellular currents were averaged across trials,
\begin{align}
\langle{I_i}\rangle = \frac{1}{10} \sum \limits_{k=1}^{10} I_{i,k}(t)
\end{align}
to remove trial-to-trial noise. We then used reverse correlation techniques to estimate the empirical frequency response function for each cell,
\begin{align}
{\hat{F}_{\mathrm{ChR2},i}}(\omega)&=\frac{S_{s\langle I_i \rangle}}{S_{ss}},\label{eq:CN1}
\end{align}
where $S_{ss}$ is the power spectrum of $s(t)$ and $S_{s\langle I_i \rangle}$ is the cross spectrum of $\langle I_i \rangle$ and $s(t)$. $S_{ss}$ and $S_{s\langle I_i \rangle}$ are defined as the Fourier transforms of the corresponding correlation function,
\begin{align}
c_{ss}(\tau)&= \int_{-T}^T \! s(t)s(t+\tau) \, \mathrm{d}\tau \label{eq:CrossCorr1} \\
c_{s\langle I_i \rangle}(\tau)&=\int_{-T}^T \! s(t) I(t+\tau) \, \mathrm{d}\tau \label{eq:CrossCorr2}.
\end{align} 
Finally, we averaged ${\hat{F}_{\mathrm{ChR2},i}}(\omega)$ across cells to obtain the empirical frequency response for each construct, ${\hat{F}_{\mathrm{ChR2}}}(\omega)$. To improve our estimate of the power spectra, we followed the procedure introduced in~\cite{Higgs} and used a frequency dependent window, equivalent to a Gaussian bandpass filter with standard deviation of $\sigma=2\pi/\omega$ in the frequency domain. Spectra were evaluated at discrete increments, $\omega_n = 2 \pi 10^n$, $n = 0.1,0.2,...,3$ . 

\section*{Results}
\label{sec:Results}
\begin{figure}[h!]
\centering
\includegraphics{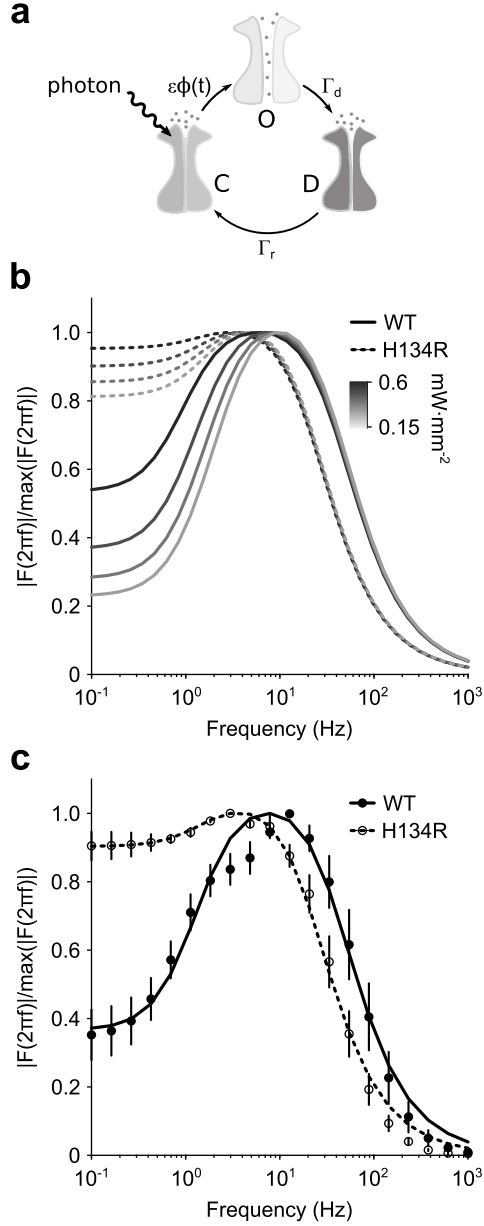}
\caption{{\bf ChR2's amplitude response function.} {\bf (a)} Illustration of the three-state Markov channel model described by \refeqn{eq:DiffEqOpenProb1}{eq:DiffEqOpenProb3}. The transition rates between open, $O$,  desensitized, $D$, and closed, $C$, states are $\epsilon\phi(t)$, $\Gamma_r$, and $\Gamma_d$. {\bf (b)} Amplitude response functions for the model, \refeq{eq:ChRResponseAmplitude}, is shown for different mean illumination intensities (0.15 through 0.6 mW$\cdot$mm$^{-2}$). {\bf (c)} Predicted amplitude response of wtChR2 (solid lines) and ChR2(H134R) (dashed lines) compared to experimentally measured response functions for a mean illumination intensity of 0.4 mW$\cdot$mm$^{-2}$. Error bars are $\pm 1$ STD ($n=9$ neurons for wtChR2 and $n=4$ neurons for ChR2(H134R)).} 
\label{fig:ResponseChRVsP}
\end{figure}

In this study, we sought a general description of ChR2's dynamics that captured the ability of both wild type ChR2 (wtChR2) and its engineered variants to convey time-varying stimuli. To do this, we determined the frequency response function of a population of channels expressed by a single cell, $F_{\mathrm{ChR2}}(\omega)$, using a Markov model of ChR2's channel kinetics~\cite{Nagel2003} (see Methods for details). The model is illustrated in Fig.~\ref{fig:ResponseChRVsP}(a) and $F_{\mathrm{ChR2}}(\omega)$ is given by \refeq{eq:ComplexR}. The current that is passed by ChR2 across the cell membrane is proportional to the number of channels that occupy the open state. Therefore, $F_{\mathrm{ChR2}}(\omega)$ can be thought of as a frequency- and phase-dependent transition rate from the channels' closed to open state in response to a time-varying stimulus. Since individual channels switch between states discretely, $F_{\mathrm{ChR2}}(\omega)$ describes the transfer of optical waveforms to intracellular current under the assumption that a large number of channels are present in the cell's membrane.

The amplitude response function, $|F_{\mathrm{wtChR2}}(\omega)|$, \refeq{eq:ChRResponseAmplitude}, indicates the frequency dependent gain of the channel population to incoming light signals. The predicted amplitude response function for wtChR2, $|F_{\mathrm{wtChR2}}(\omega)|$, and the H134R mutant, $|F_{\mathrm{ChR2(H134R)}}(\omega)|$, are shown in Fig.~\ref{fig:ResponseChRVsP}(b) for different mean illumination intensities. $|F_{\mathrm{wtChR2}}(\omega)|$ has high frequency cutoff (full width at half maximum) of 69 Hz. It should be noted that this cutoff value is defined relative to wtChR2's peak conductance, and not in terms of absolute photocurrents. For this reason, it is possible to use wtChR2 to deliver physiologically significant photocurrents at stimulation frequencies exceeding 69 Hz. The shape of $|F_{\mathrm{wtChR2}}(\omega)|$ indicates that wtChR2 exhibits a significant resonance with a natural frequency close to 10 Hz. This feature explains the large peak to steady-state (DC) current ratio exhibited by wtChR2~\cite{Mattis2011}. In agreement with previous characterizations of the channel, the H134R variant is significantly slower than wtChR2 and $|F_{\mathrm{ChR2(H134R)}}(\omega)|$ has a cutoff frequency at 37 Hz. While ChR2(H134R) supports a resonance, it is significantly reduced compared to wtChR2.

\begin{figure}[h!]
\centering
\includegraphics{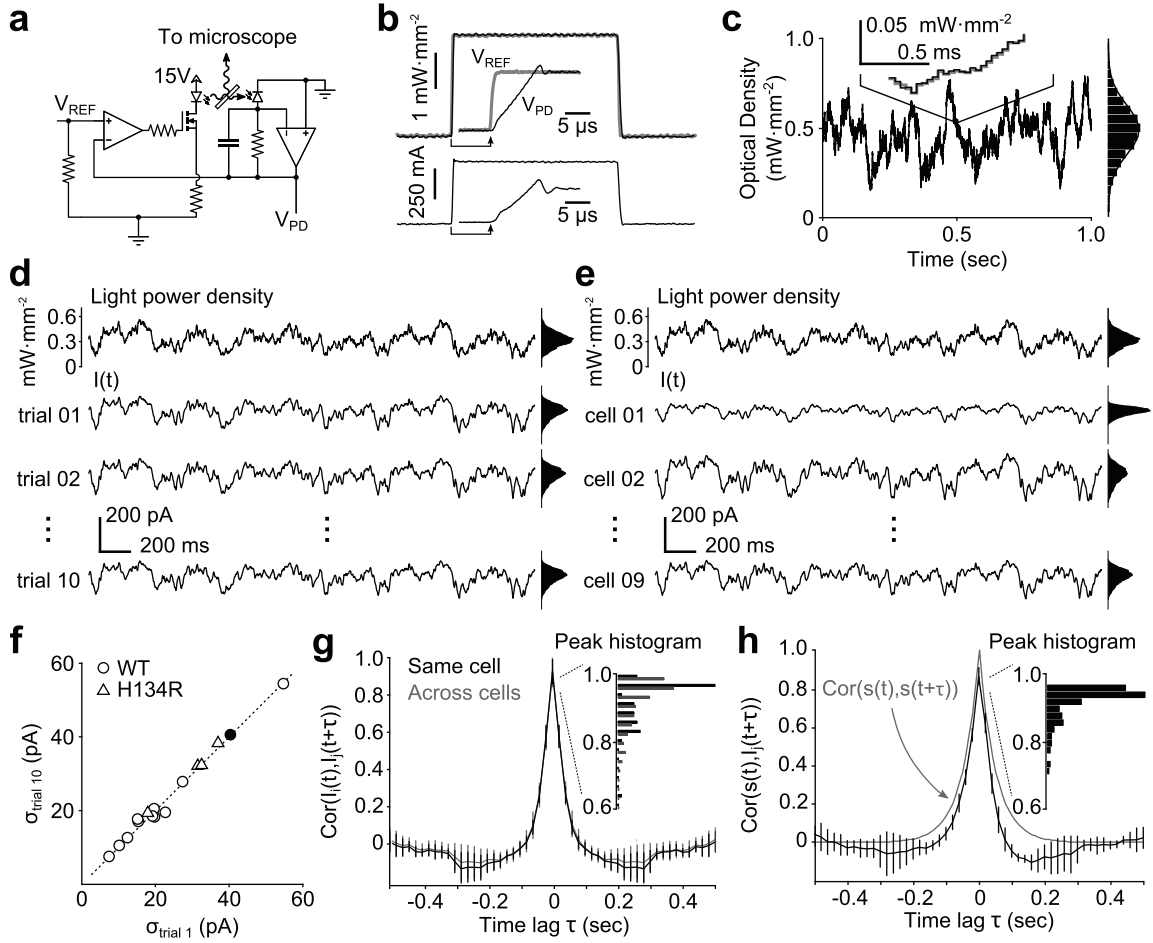}
\caption{{\bf Reliability of time-varying photostimulation.} {\bf (a)} Simplified schematic of LED driver in optical feedback mode. {\bf (b)} One millisecond LED pulse, ${\rm V_{PD}}$ (black), versus the reference voltage, ${\rm V_{REF}}$ (gray). Inset shows the step onset with corresponding scale bars. The current sourced to the LED is shown in the bottom plot. {\bf (c)} A DAC-controlled Gaussian stimulus (gray) signal and recorded light waveform (black). Inset shows zoomed portion of the sequence. An amplitude histogram of the sequence, with best-fit Gaussian, is shown to the right. {\bf (d)} Intracellular currents, from a single cell, during Gaussian stimuli. The top trace is a portion of a 10 second Gaussian stimulus sequence. The bottom three traces show the intracellular currents recorded during different presentations of the same stimulus waveform to a single cell. {\bf (e)} The same stimulus waveform used in (d), and the corresponding evoked responses from different cells. {\bf (f)} The standard deviation of the photocurrent induced on the first trial of stimulation versus the last trial. The dotted line is identity. The filled dot corresponds to the cell in (d). {\bf (g)} Cross/auto-correlation functions of photocurrents between neurons, $I_i(t)$. The inset histogram shows the peak correlation coefficients. {\bf (h)} Cross-correlation function between the stimulation process $s(t)$ and recorded photocurrents. The grey line is the autocorrelation function of the stimulation process. The inset histogram shows the peak correlation coefficients. Unless otherwise noted, data in this figure were obtained from cells expressing wtChR2.}
\label{fig:Reliability}
\end{figure}

To verify $F_{\mathrm{ChR2}}(\omega)$ experimentally, cultured cortical cells expressing either wtChR2 or ChR2(H134R) were stimulated with 465 nm at peak intensity, spatially uniform light while somatic photocurrents were recorded using whole cell patch clamp (see Methods for details). Fig.~\ref{fig:Reliability}(a-c) details the stimulation system used in our study. Optical stimuli consisted of a 10 second realization of a Gaussian noise process with a time constant of 50 ms and a mean$\pm$STD  irradiance of $0.4 \pm 0.08 \mathrm{mW}\mathrm{mm}^{-2}$.  We choose stimuli with these parameters because they evoked currents with similar amplitude and temporal characteristics to those obtained from {\em in vivo} recordings of sensory cortical neurons~\cite{Hestrin,Stern1992,destexhe2003}. 

We estimated the empirical frequency response function, $\hat{F}_\mathrm{ChR2}(\omega)$, for cells expressing wtChR2 ($n = 9$ cells) or ChR2(H134R) ($n = 4$ cells) using reverse correlation techniques (\refeq{eq:CN1}). Fig.~\ref{fig:ResponseChRVsP}(c) compares the empirical amplitude response, $|\hat{F}_\mathrm{ChR2}(\omega)|$, of wtChR2 and ChR2(H134R) and their theoretical counterparts. We observed good agreement between the predicted response function and $|\hat{F}_\mathrm{ChR2}(\omega)|$, although some differences exist. For instance, both $|F_{\mathrm{ChR2(H134R)}}(\omega)|$ and $|F_{\mathrm{wtChR2}}(\omega)|$ have a small downward deviation from the predicted response at $\sim$5 Hz, which is more pronounced for wtChR2. Additionally, the predicted frequency response tends to slightly overestimate the measured gain at frequencies above 100 Hz. Despite these differences, both theoretical and empirical results indicate that wtChR2 and ChR2(H134R) are useful tools for transmitting fluctuating current stimuli to populations of cells in a physiologically relevant frequency range. Additionally, because the model provides an extensible description of channel dynamics, it serves as a useful tool for predicting the bandwidth and resonance of new channels based on measurable physiological parameters.

Next, we measured the reliability of photocurrents across trials. Fig.~\ref{fig:Reliability}(d) shows a 3-second portion of a stimulus waveform and the corresponding photocurrents for a single cell expressing wtChR2, across trials. As expected, these photocurrents appear to be smoothed versions of the stimulus signal due to the low-pass effect of wtChR2's amplitude response function. Photocurrents were remarkably stable across trials. The trial-to-trial repeatability of evoked photocurrent waveforms is captured in Fig.~\ref{fig:Reliability}(f), which displays the standard deviation of the current waveform recorded during the first presentation of the stimulus compared to the 10th presentation, for each cell. Because these points fall near the identity line, it can be inferred that there is no change in the efficacy of photostimulation across trials.

We then examined the distribution of evoked current amplitudes across cells ( Fig.~\ref{fig:Reliability}(f)). wtChR2 delivered current waveforms with a mean waveform standard deviation of 26.7 pA per cell. ChR2(H134R) delivered only slightly larger current fluctuations than wtChR2, with a mean waveform standard deviation of 32.0 pA per cell. The comparable amplitudes of evoked currents between wtChR2 and ChR2(H134R) is due to wtChR2's strong resonance, which makes it more sensitive to fluctuating, compared to DC, optical waveforms.  

Finally, to determine the reliability of evoked currents across cells, we calculated the normalized cross-correlation function, $c_{sI_i}$, between the light power density, $s(t)$ and photocurrents $I_i(t)$ at each cell, $i$, and across cells, $c_{I_i, I_j}$.  Fig.~\ref{fig:ResponseChRVsP} (g) and (h) show  $c_{sI_i}$ and $c_{I_i, I_j}$ for wtChR2. The median peak value of $c_{I_iI_j}$ is $0.96$, indicating strong correlations between evoked currents in different cells. The median peak value of $c_{sI_i}$ is $0.92$, indicating strong correlations between evoked currents and the stimulus waveform. Additionally, the similarity in shape between $c_{sI}$ and the autocorrelation of the stimulation process, $c_{ss}$, indicates that precise temporal features of the stimulus were accurately converted into a photocurrent, as predicted by the passband of the frequency response functions.

\section*{Discussion}
\label{sec:Discussion}
Optogenetic methods offer genetic specificity, elimination of electrical recording artifacts, and an expanding toolset with increasingly specialized functionality~\cite{Mattis2011}. Because of these advantages, optogenetic methods are often used for direct interaction with neuronal subpopulations in order to understand their function~\cite{Cardin2009,Sohal2009a}. Typically, pulsed optical stimuli are used to probe neural circuits. However, for studies that seek to understand information transmission in neural circuits, continuously modulated photocurrents better mimic natural synaptic bombardment than do pulsed stimuli.

In this study, we demonstrated the ability of ChR2 to deliver continuously modulated photocurrents to neurons. We used a three state Markov model~\cite{Nagel2003} to derived an analytical frequency response function for ChR2 variants (\refeq{eq:ComplexR}). We confirmed this model  experimentally indicating that \refeq{eq:ComplexR} is sufficient to capture dynamical properties of ChR2 in neurons. Additionally, we found that the passband of wtChR2 and ChR2(H134R) is broad enough to support photocurrents that mimic natural synaptic communication. The reliability of somatic photocurrents is very high, with a correlation strength upward of $0.9$ across trials and cells. ChR2's amplitude response function, \refeq{eq:ChRResponseAmplitude}, indicates that the sum of channel recovery and desensitization transition rates determines the frequency cut-off. Therefore, opsins with faster transition rates~\cite{Gunaydin2010} will likely allow an even broader dynamic range. Interestingly, faster channels and increased bandwidth may eventually offer an artificial, optical neural communication channel that actually exceeds the bandwidth of natural sensory organs. This would have tremendous implications for how neural computation and processing are studied and for the advancement of brain-machine interfaces. Finally, we showed that ChR2's frequency response function supports a resonance. The degree of resonance is dependent on the values of free model parameters, which will change for different ChR2 variants. This finding is especially important for studies that use ChR2 to examine the frequency-dependence of neural circuitry~\cite{Cardin2009}, since it is important to separate the intrinsic dynamics of ChR2 from those that belong to the network under study.  

ChR2 was derived from microbes that use it for optical sensation in natural environments. It is therefore not surprising that the channel is excellent at conveying time-varying input signals. Using wtChR2 and its numerous variants as a means for delivering repeatable, time-varying stimuli to genetically defined populations of cells will be a powerful method for probing the dynamics of neural circuits.

\bibliography{ou-chr2}

\begin{thebibliography}{10}
\providecommand{\url}[1]{\texttt{#1}}
\providecommand{\urlprefix}{URL }
\expandafter\ifx\csname urlstyle\endcsname\relax
  \providecommand{\doi}[1]{doi:\discretionary{}{}{}#1}\else
  \providecommand{\doi}{doi:\discretionary{}{}{}\begingroup
  \urlstyle{rm}\Url}\fi
\providecommand{\bibAnnoteFile}[1]{%
  \IfFileExists{#1}{\begin{quotation}\noindent\textsc{Key:} #1\\
  \textsc{Annotation:}\ \input{#1}\end{quotation}}{}}
\providecommand{\bibAnnote}[2]{%
  \begin{quotation}\noindent\textsc{Key:} #1\\
  \textsc{Annotation:}\ #2\end{quotation}}
\providecommand{\eprint}[2][]{\url{#2}}

\bibitem{Warland1997}
Warland DK, Reinagel P, Meister M (1997) {Decoding visual information from a
  population of retinal ganglion cells}.
\newblock J Neurophysiol 78: 2336--50.
\bibAnnoteFile{Warland1997}

\bibitem{Chichilnisky2001}
Chichilnisky EJ (2001) A simple white noise analysis of neuronal light
  responses.
\newblock Network 12: 199-213.
\bibAnnoteFile{Chichilnisky2001}

\bibitem{Geffen2009}
Geffen MN, Broome BM, Laurent G, Meister M (2009) {Neural encoding of rapidly
  fluctuating odors.}
\newblock Neuron 61: 570-86.
\bibAnnoteFile{Geffen2009}

\bibitem{Mattis2011}
Mattis J, Tye K, Ferenczi E, Ramakrishnan C, O'Shea D, et~al. (2011)
  {Principles for applying optogenetic tools derived from direct comparative
  analysis of microbial opsins}.
\newblock Nature Methods 99:2: 159--172.
\bibAnnoteFile{Mattis2011}

\bibitem{Gunaydin2010}
Gunaydin LA, Yizhar O, Berndt A, Sohal VS, Deisseroth K, et~al. (2010)
  {Ultrafast optogenetic control.}
\newblock Nature Neuroscience 13: 387--92.
\bibAnnoteFile{Gunaydin2010}

\bibitem{destexhe2003}
Destexhe A, Rudolph M, Pare D (2003) The high-conductance state of neocortical
  neurons {\em in vivo}.
\newblock {Nat Rev Neurosci} 4: 739-751.
\bibAnnoteFile{destexhe2003}

\bibitem{Mainen1995}
Mainen ZF, Sejnowski TJ (1995) Reliability of spike timing in neocortical
  neurons.
\newblock Science 268: 1503-1506.
\bibAnnoteFile{Mainen1995}

\bibitem{TchumJN}
Tchumatchenko T, Malyshev A, Wolf F, Volgushev M (2011) Ultra-fast population
  encoding by cortical neurons.
\newblock J Neurosci 31: 12171-12179.
\bibAnnoteFile{TchumJN}

\bibitem{Boyden2005}
Boyden ES, Zhang F, Bamberg E, Nagel G, Deisseroth K (2005)
  {Millisecond-timescale, genetically targeted optical control of neural
  activity.}
\newblock Nature neuroscience 8: 1263--8.
\bibAnnoteFile{Boyden2005}

\bibitem{Nagel2005}
Nagel G, Brauner M, Liewald JF, Adeishvili N, Bamberg E, et~al. (2005) {Light
  activation of channelrhodopsin-2 in excitable cells of Caenorhabditis elegans
  triggers rapid behavioral responses.}
\newblock Current biology 15: 2279--84.
\bibAnnoteFile{Nagel2005}

\bibitem{Nagel2003}
Nagel G, Szellas T, Huhn W, Kateriya S, Adeishvili N, et~al. (2003)
  {Channelrhodopsin-2, a directly light-gated cation-selective membrane
  channel.}
\newblock PNAS 100: 13940--5.
\bibAnnoteFile{Nagel2003}

\bibitem{Hales2010}
Hales CM, Rolston JD, Potter SM (2010) {How to culture, record and stimulate
  neuronal networks on micro-electrode arrays}.
\newblock J Visualized Exp 39: e2056.
\bibAnnoteFile{Hales2010}

\bibitem{Potter2001}
Potter SM, DeMarse TB (2001) {A new approach to neural cell culture for
  long-term studies}.
\newblock J Neurosci Methods 110: 17--24.
\bibAnnoteFile{Potter2001}

\bibitem{Higgs}
Higgs MH, Spain WJ (2009) Conditional bursting enhances resonant firing in
  neocortical layer 2-3 pyramidal neurons.
\newblock J Neurosci 29: 1285--1299.
\bibAnnoteFile{Higgs}

\bibitem{Hestrin}
Hestrin S (1993) Different glutamate receptor channels mediate fast excitatory
  synaptic currents in inhibitory and excitatory cortical neurons.
\newblock Neuron 11: 1083-1091.
\bibAnnoteFile{Hestrin}

\bibitem{Stern1992}
Stern P, Edwards FA, Sakmann B (1992) {Fast and slow components of unitary
  EPSCs on stellate cells elicited by focal stimulation in slices of rat visual
  cortex}.
\newblock Physiology 449: 247--278.
\bibAnnoteFile{Stern1992}

\bibitem{Cardin2009}
Cardin J, Carlen M, Meletis K, Knoblich U, Zhang F, et~al. (2009) Driving
  fast-spiking cells induces gamma rhythm and controls sensory responses.
\newblock Nature 459: 663--667.
\bibAnnoteFile{Cardin2009}

\bibitem{Sohal2009a}
Sohal V, Zhang F, Yizhar O, Deisseroth K (2009) {Parvalbumin neurons and gamma
  rhythms enhance cortical circuit performance}.
\newblock Nature 459: 698--702.
\bibAnnoteFile{Sohal2009a}

\end{thebibliography}
\bibliographystyle{plos2009}

\newpage
\section*{Figures}

\end{document}